\documentclass[12pt]{article}
\usepackage{amsmath}
\usepackage{amsfonts}
\usepackage{amssymb}
\usepackage{latexsym}
\usepackage{graphicx}
\usepackage[english]{babel}
\input epsf.sty

\begin{document}

\topmargin 0pt
\oddsidemargin 0mm



\begin{flushright}

\hfill{USTC-ICTS-06-10}\\

\end{flushright}
\vspace{4mm}

\begin{center}

{\Large \bf

Inflation with High Derivative Couplings}

\vspace{8mm}

{\large Bin Chen$^{1}$, Miao Li$^{2,3}$, Tower Wang$^{2}$, Yi Wang$^{2}$}

\vspace{4mm}

{\em
$^{1}$ School of Physics, Peking University, Beijing 100871, P.R.China\\
 $^{2}$ Institute of Theoretical Physics, Academia Sinica, Beijing 100080, P.R.China\\
 $^{3}$ The Interdisciplinary Center for Theoretical Study\\}

\end{center}

\vspace{6mm}
We study a class of generalized inflation models in which the
inflaton is coupled to the Ricci scalar by a general $f(\varphi, R)$
term. The scalar power spectrum, the spectral index, the running of the
spectral index, the tensor mode spectrum and a new consistency
relation of the model are calculated. We discuss in detail the
issues of how to diagonize the coupled perturbation equations, how
to deal with an entropy-like source, and how to determine the
initial condition by quantization. By studying some explicit models,
we find that rich phenomena such as a blue scalar power spectrum, a
large running of the spectral index, and a blue tensor mode spectrum
can be obtained.

\newpage


\section{Introduction}

Inflation \cite{inflation} has been very successful in solving the
problems in the standard big bang cosmology such as the horizon
and flatness problems. Fluctuations are created during inflation,
which provide seeds of large scale structure and the Cosmic
Microwave Background (CMB) anisotropies having been observed by
COBE and WMAP satellites, the latter will be measured more precisely by
PLANCK.

In spite of the remarkable achievements, there are still open
problems, such as a natural realization of inflation models in a
fundamental theory, and how to further pin down a specific model
including the number of the inflatons and the form of the potential.
In ref.\cite{Bt0509099}, some theoretical problems are listed. For
example, the hierarchy of scales, the trans-Planck problem, the
singularity problem and the candidate of the inflaton. As for the
data fitting, there is still a significant power deficit at $l=2$,
and if the running of the spectral index is introduced as a
parameter, the WMAP three year data mildly favors a blue spectrum
with a large running \cite{largerunning}, which is not common in
inflation models. In the past, a few models were proposed to
accommodate a large negative running such as the noncommutative
inflation model \cite{noncomm}. For other models we refer to
\cite{owlarge1,owlarge2,owlarge3,owlarge4}. Discussions on models
violating the null energy condition and leading to a blue tensor spectrum
and possibly a blue scalar spectrum, as well as on the WMAP three year constraints,
can be found in \cite{bfm}

To have a broad scope, and to be prepared for possible surprises in
the future experiments, it is essential to study classes of models
with novel physics built in, for example, quantum gravity and string
theory may induce higher derivative corrections. In this paper, our
primary goal is to carry out  a more careful study of a class of
models in which in addition to the Hilbert-Einstein action and a
minimal coupled inflaton, there is a general nominimal coupling of
the form $f(\varphi, R)$. Some of these models were studied before.
However, when we examined the literature carefully, we found that
only some special models were reliably studied. These include:
$f(R)$ gravity models \cite{fR} and the $f(\varphi) R$ gravity
models \cite{fphiR}, a unified analysis of these two cases were made
in \cite{H0412}.
 Furthermore, even for  $f(R)$ action and
$f(\varphi) R$ action, no concrete models were investigated in
detail. Remarkably, the result for the tensor modes power spectrum
in \cite{H0412} is applicable to more general cases. We note, however,
that the pioneering work on inflation with an additional scalar curvature squared
term was done long time ago \cite{alexei}

In \cite{miaoli}, the models with a nonminimal coupling term
$-\frac{1}{2}f(\frac{1}{6}R)\varphi^2$ are studied, and the
calculations are carried out when $f$ is quadratic in $R$ and the
nonminimal coupling gives dominate contribution in the rolling of
$\varphi$. It is shown that a blue spectrum with a running larger
than that of a common minimal model can be produced. However, the
constraints on the fluctuation of the inflaton and the scalar
fluctuations in the metric are over-simplified in \cite{miaoli}.

In this paper, we study a general class of models in which the
coupling between the inflaton $\varphi$ and the Ricci scalar $R$ is
described by a general function $f(\varphi, R)$. Both the background
dynamics and the perturbations are studied, and the power spectrum,
the spectral index, the running of the spectral index and the
perturbations for the tensor modes are calculated, which can be
related to the experiments today and in the future.

Some ingredients in the calculations are standard as being used in
the minimal inflation models (\cite{MFB92} is a nice review). Due
to the higher derivative terms, the constraints among the
fluctuations are no longer purely algebraic. We develop a general
method to deal with this problem. Diagonalization of the
perturbation equations is no longer simple. Upon assuming a
slow-roll inflation, we can still have a nearly-conserved quantity
which reduces to the usual curvature perturbation when the higher
derivative terms are suppressed.
 Thus, although there is
an entropy-like perturbation, we can bypass this problem by using
this new nearly-conserved quantity. Also, since we can not
directly write down the action for the perturbation modes, we
provide an argument in relating the perturbation in a diagonalized
perturbation equation to the canonically normalized mode, thus set
the initial quantization condition. Some of these methods are
quite general and can be used in other generalizations of the
minimal inflation models.

By calculating the power spectrum and related quantities, we find
that there are rich phenomena in this class of models. We can have
a blue spectrum and a large running of the spectral index in some
models, while in other models the spectrum stay red and the
running can become positive. Also, in principle, one can have an
initial fast-rolling period when the higher derivative terms are
comparable to the Hilbert-Einstein term. The consistency condition
between $r$ and $n_T$ is modified.  A blue tensor mode power
spectrum can appear even in some small inflaton models in some
parameter region while in minimal inflation models the tensor
spectrum is always red.

The paper is organized as follows. In Section 2, we derive the
equations of motion of background quantities, and spell out the
conditions of slow-roll approximation. In Section 3, we write down
the equations for perturbations. After diagonalizing and simplifying
the equations to obtain a Bessel equation, we propose a
nearly-conserved quantity which has the form of comoving curvature
perturbation in the minimal inflation model limit. In Section 4, we
study the initial conditions by relating the fluctuation under study
to the canonically normalized mode. Since the canonical variables
are normalized by quantization, this procedure help us to fix the
integration constants in the solutions of the Bessel equation. In
Section 5, the power spectrum, the spectral index, the running of the
spectral index, and the ratio of the tensor and scalar perturbations
are given. In Section 6, we calculate several classes of models and
discuss the properties of these models. We conclude in  Section 7.

\section{The Model and Background Dynamics}

We consider the following action with nonminimal coupling:
\begin{equation}
S=\int
d^4x\sqrt{-g}[\frac{1}{2}f(\varphi,R)-\frac{1}{2}g^{\alpha\beta}\partial_{\alpha}\varphi\partial_{\beta}
\varphi-V(\varphi)],\label{lagrangian}
\end{equation}

We  consider first the evolution of the background,  a
homogeneous and isotropic spacetime with a space-independent
inflation field. We work with the
Friedmann-Robertson-Walker(FRW) metric
\begin{equation}
ds^2=-dt^2+a^2(t)dx^2=a^2(\tau)(-d\tau^2+dx^2),
\end{equation}
where $t$ is the comoving time and $\tau$ is the conformal time.

It follows from the action (\ref{lagrangian}) that, the equations of
motion are
\begin{eqnarray}
&&H^2=\frac{1}{3M_p^{2}}\rho,~~\dot{H}=-\frac{1}{2M_p^{2}}(\rho+p),\\
&&\ddot{\varphi}+3H\dot{\varphi}+\frac{1}{2}(-f_{, \varphi}+2V_{, \varphi})=0,
\end{eqnarray}
where $\rho$ and $p$ are energy density and pressure respectively, as the diagonal
elements of the spacial averaged energy-momentum tensor, and can
be written as
\begin{eqnarray}
\rho &=&\frac{M_p^{2}}{F}(\frac{1}{2}\dot{\varphi}^2+V+\frac{RF-f}{2}-3H\dot{F}),\\
p&=&\frac{M_p^{2}}{F}(\frac{1}{2}\dot{\varphi}^2-V-\frac{RF-f}{2}+\ddot{F}+2H\dot{F}),
\end{eqnarray}
where $F=\frac{\partial}{\partial R}f(\varphi,R)$, and dot denotes the
derivative with respect to $t$.

We shall impose the following slow-roll conditions:
\begin{eqnarray}
\epsilon&\equiv&-\frac{\dot{H}}{H^2}\ll 1,\\
\delta&\equiv&-\frac{\ddot{\varphi}}{H\dot\varphi}\ll 1,\\
\gamma&\equiv&\frac{\dot{F}}{H F}\ll 1, \label{slowroll}
\end{eqnarray}
The first and second slow-roll conditions are standard as in the
minimal inflation model. The third slow-roll condition is new in the
nonminimal model. It is a natural requirement, because $F$ can be
written as a function of $R$ and $\varphi$ and should be
slow-rolling during inflation. Moreover, it will be shown that
$\gamma$ appears in the spectral index and the ratio of tensor and
scalar perturbations. To be consistent with the CMB data,  it should
be small in any viable candidate inflation model. In addition to the
above slow-roll conditions, we further require that the above
slow-roll parameters themselves are rolling slowly.

\section{Classical Evolution of Perturbations}

To calculate the scalar fluctuations, we need to perturb the
Einstein equations. Through out the calculations, we will work in
the longitudinal gauge,
\begin{equation}
ds^2=a^2\left(-(1+2\phi)d\tau^2+(1-2\psi)dx^idx^i\right),
\end{equation}
and the fluctuation of the inflaton $\varphi$ is denoted by
$\delta \varphi$.

The perturbation equations are
\begin{eqnarray}
\label{g00}&&3H(H\phi+\dot{\psi})-\frac{1}{a^2}\nabla^2\psi=-\frac{1}{2 M_p^{2}}\delta\rho,\\
\label{g0j}&&3(H\phi+\dot{\psi})=\frac{3}{2 M_p^{2}}(\rho+p)v,\\
\label{gij}&&\psi-\phi=\frac{\delta F}{F},\\
\label{gii}&&\ddot{\psi}+3H(H\phi+\dot{\psi})+H\dot{\phi}+2\dot{H}\phi+\frac{1}{3a^2}\nabla^2(\phi-\psi)=\frac{1}{2
M_p^{2}}\delta p,
\end{eqnarray}
which follow from the components $G^0_0, G^0_j, G^i_j(i\neq j)$,
$G^i_i$ of the linearized Einstein equations for perturbations,
respectively. The $\delta\rho$, $\delta p$ and $v$ are
perturbative quantities in the energy momentum tensor defined as,
\begin{eqnarray}
T^0_0=-(\rho+\delta\rho),\hspace{3ex}T^0_i=-(\rho+p)v_{,i},\hspace{3ex}T^i_i=3(p+\delta
p),
\end{eqnarray}
with the explicit form
\begin{eqnarray}
\delta\rho&=&\frac{M_p^{2}}{F}[\dot{\varphi}\delta\dot{\varphi}+\frac{1}{2}(-f_{,\varphi}+2V_{,\varphi})\delta\varphi-3H\delta\dot{F}+(3\dot{H}+3H^2+\frac{1}{a^2}\nabla^2)\delta F\nonumber\\
&&+(3H\dot{F}-\dot{\varphi}^2)\phi+3\dot{F}(H\phi+\dot{\psi})],\\
\delta p&=&\frac{M_p^{2}}{F}[\dot{\varphi}\delta\dot{\varphi}+\frac{1}{2}(f_{,\varphi}-2V_{,\varphi})\delta\varphi+\delta\ddot{F}+2H\delta\dot{F}+(-\dot{H}-3H^2-\frac{2}{3a^2}\nabla^2)\delta F\nonumber\\
&&-\dot{F}\dot{\phi}-(\dot{\varphi}^2+2\ddot{F}+2H\dot{F})\phi-2\dot{F}(H\phi+\dot{\psi})],\\
(\rho+p)v&=&\frac{M_p^{2}}{F}(\dot{\varphi}\delta\varphi+\delta\dot{F}-H\delta
F-\dot{F}\phi)
\end{eqnarray}

By using equations (\ref{g0j}) and (\ref{gij}) to cancel $\delta
\varphi$ and $\delta F$ terms in (\ref{g00}) respectively, one
gets
\begin{eqnarray}
\nonumber&&F(\ddot{\phi}+\ddot{\psi})+(HF+3\dot{F}-\frac{2F\ddot{\varphi}}{\dot{\varphi}})(\dot{\phi}+\dot{\psi})\\
\nonumber &&+[(3H\dot{F}+3\ddot{F})-\frac{2\ddot{\varphi}}{\dot{\varphi}}(HF+2\dot{F})-\frac{F}{a^2}\nabla^2]\phi\\
&&+[(4\dot{H}F+H\dot{F}-\ddot{F})-\frac{2\ddot{\varphi}}{\dot{\varphi}}(HF-\dot{F})-\frac{F}{a^2}\nabla^2]\psi=0
\label{purt1}
\end{eqnarray}

And by using $\delta F=F_{,R}\delta R+F_{,\varphi}\delta\varphi$ and
$\dot{F}=F_{,R}\dot{R}+F_{,\varphi}\dot{\varphi}$, we can write
(\ref{gij}) in a more explicit form
\begin{eqnarray}
&&F(2\dot{H}F-H\dot{F}+\ddot{F})(\phi-\psi)\nonumber\\
\nonumber &&+[6F_{,R}(4H\dot{H}+\ddot{H})-\dot{F}][F(\dot{\phi}+\dot{\psi})+(HF+2\dot{F})\phi+(HF-\dot{F})\psi]=\\
&&2F_{,R}(2\dot{H}F-H\dot{F}+\ddot{F})\times\nonumber\\
&&[3\ddot{\psi}+3H\dot{\phi}+12H\dot{\psi}+6(2H^2+\dot{H})\phi+\frac{1}{a^2}\nabla^2(\phi-2\psi)]
\label{purt2}
\end{eqnarray}
In the limit of minimal inflation model, (\ref{purt2}) reduces to
the constraint equation $\psi=\phi$.

Equations (\ref{purt1}) and (\ref{purt2}) are coupled
differential equations for $\psi$ and $\phi$. In some special
cases, for example, $F=F(\varphi)$ or $F=F(R)$, these equations can
be diagonalized to a single differential equation \cite{H0412}.
While in a more general case which is of main interest in this paper,
we have to develop a new method to solve these equations.

To diagonize the equations, we define
\begin{equation}
{\cal G}_k\equiv \psi_k/\phi_k,\label{const}
\end{equation}
where  $\psi_k$ and $\phi_k$ are Fourier components of $\psi$ and
$\phi$, and the explicit $k$-dependence of ${\cal G}_k$ will be
shown later in this section. We can Taylor expand ${\cal G}_k$ as
\begin{equation}
{\cal G}_k=G_{0}+\frac{F_{,R} H^2}{F}\left(\frac{k}{aH}\right)^2
G_{1}+\cdots,\label{planewave}
\end{equation}
where $G_0$ and $G_1$ stand for the zeroth and first coefficients
in the Taylor expansion.

So under the condition that the nonminimal coupling is not too
strong compared with the minimal coupling gravity, the above
expansion works well near the horizon-crossing and we will see in
the following sections that $G_{0}$ will be the dominant
contribution. It ought to be true for physically interesting
models where the energy scale of inflation is smaller than the
Planck scale or the string scale. While when the expansion does
not work well and the high order terms of $k$ become important,
the high order derivative corrections will play a much more
important role at the Hubble scale. In this case the dispersion
relation will be modified significantly when the perturbation was
created, and we need to modify the usually canonical quantization.
A quantum gravity theory perhaps is in demand in order to fully
understand this regime.

We now derive ${\cal G}_k$ as a function of background quantities
from  (\ref{purt1}) and (\ref{purt2}). Substitute (\ref{const})
into (\ref{purt1}) and (\ref{purt2}) to eliminate $\psi_k$, then
use (\ref{purt1}) and (\ref{purt2}) to eliminate the $\phi{''}$
terms to get a first-order differential equation. Differentiate
this equation once and subtract with (\ref{purt1}), we can get
another first-order differential equation. From the consistence
condition of the two first-order differential equations, we can
get the expression of ${\cal G}_k$, which to the lowest order is
the solution of the following equation:

\begin{eqnarray}
\nonumber &&\{[F\dot{F}+24HF_{,R}(\dot{H}F-H\dot{F})]{\cal G}_k+F(\dot{F}-24H\dot{H}F_{,R})\}\\
\nonumber &&\times\{[2H\dot{H}F(F-12H^2F_{,R})-4HF_{,R}(4\dot{H}F-5H\dot{F})\frac{k^2}{a^2}-F\dot{F}\frac{k^2}{a^2}]{\cal G}_k\\
\nonumber &&-2H(F-12H^2F_{,R})(\dot{H}F-H\dot{F})+8HF_{,R}(\dot{H}F+H\dot{F})\frac{k^2}{a^2}-F\dot{F}\frac{k^2}{a^2}\}\\
\nonumber &=&\{[2\dot{H}F(F-12H^2F_{,R})-2F_{,R}(2\dot{H}F-H\dot{F})\frac{k^2}{a^2}]{\cal G}_k\\
&&-2(F-12H^2F_{,R})(\dot{H}F-H\dot{F})-2F_{,R}(2\dot{H}F-H\dot{F})\frac{k^2}{a^2}\}^2,
\label{Geq}
\end{eqnarray}
we see that indeed ${\cal G}_k$ is a function of only the
background. By solving $G_{0}$ and $G_{1}$ from this equation, it
can be shown that $G_{0}$ and $G_{1}$ are slow-roll quantities and
$G_{0}$ has the simple form
\begin{equation}
G_{0}=1-\frac{\dot{F}H}{F\dot{H}}
\end{equation}

With the use of the  linear relation between $\psi_k$ and $\phi_k$,
to leading order of slow-roll parameters, we derive a differential
equation for a single variable, say $\phi_k$
\begin{eqnarray}
&&\phi{''}_k+(aH)\left(\frac{2\dot{\cal G}_k}{H(1+{\cal G}_k)}+2\delta+3\gamma\right)\phi'_k+k^2\phi_k \nonumber\\
&&+(aH)^2\left(\frac{\ddot{{\cal G}_k}}{H^2(1+{\cal
G}_k)}+\frac{\dot{{\cal G}_k}}{H(1+{\cal G}_k)}-\frac{4{\cal
G}_k}{1+{\cal G}_k}\epsilon +\frac{3+{\cal G}_k}{1+{\cal
G}_k}\gamma+2\delta \right)\phi_k=0 \label{phidiag}
\end{eqnarray}

By introducing a new variable
\begin{equation}
u_k\equiv\frac{F^{\frac{3}{2}}(1+{\cal G}_k)}{\dot{\varphi}}
\phi_k ,
\end{equation}
the equation (\ref{phidiag}) can be rewritten as
\begin{equation}
u{''}_k+\frac{1}{\tau^2}\left(\delta+\frac{3-{\cal G}_k}{2(1+{\cal
G}_k)}\gamma-\frac{4{\cal G}_k}{1+{\cal G}_k}\epsilon
\right)u_k+k^2 u_k=0,
\end{equation}
where $(aH)^2$ has been replaced by $1/\tau^2$ up to the leading
order precision.

Inserting the Taylor expansion of ${\cal G}_k$, one  finds
\begin{equation}
u{''}_k+\frac{1}{\tau^2}\left(\delta+\frac{3-G_0}{2(1+G_0)}\gamma-\frac{4G_0}{1+G_0}\epsilon
\right)u_k+\tilde{k}^2 u_k=0 \label{momentum1},
\end{equation}
where $\tilde{k}$ is a rescaling of $k$, due to the contribution
of the $k$-dependent part of ${\cal G}_k$,
\begin{equation}
\frac{\tilde{k}^2}{k^2}=1+\left(\frac{4G_0}{2(1+G_0)^2}\gamma
-\frac{4(1+2G_0)}{(1+G_0)^2}\epsilon\right)\frac{F_{,R}
H^2}{F}G_1\label{rescale}
\end{equation}
It can be shown in the models under consideration and to the
desired precision, this rescaling can be neglected, and we shall
use $k$ instead of $\tilde{k}$. If one wish to calculate to second
order in slow-roll parameters in this nonminimal model, this
rescaling should be taken into account. This effect may also arise
in models other than (\ref{lagrangian}).

The equation (\ref{momentum1}) can be further simplified to
\begin{equation}
u{''}_k+\frac{1}{2\tau^2}\left(-4\epsilon +2\delta-\gamma
\right)u_k+\tilde{k}^2 u_k=0 \label{momentum},
\end{equation}

Which takes the form of a Bessel equation. The solution of
(\ref{momentum}) can be written as
\begin{equation}\label{usolution}
u_k=C_1 \sqrt{-\tau} H^{(1)}_{\nu}(-k\tau)+C_2 \sqrt{-\tau}
H^{(2)}_{\nu}(-k\tau),
\end{equation}
where $C_1$ and $C_2$ are the integration constants to be determined by
the initial conditions, and $\nu$ is
\begin{equation}
\nu\equiv\frac{1}{2}\sqrt{1-2\left( -4\epsilon + 2\delta
-\gamma\right)}\simeq \frac{1}{2}-\frac{1}{2}(-4\epsilon
+2\delta-\gamma )\label{nu}
\end{equation}

There are rich phenomenona as preheating and reheating for the
perturbations from crossing the horizon to reentering the horizon.
In order not to track the whole evolution process, we need to find a
conserved quantity well outside the horizon. In the minimal
inflation models, the comoving curvature perturbation is introduced
to deal with this problem. But in the models with entropy
perturbation, the comoving curvature perturbation and its direct
generalizations may not be conserved. In this paper, we develop a
general method to derive a nearly conserved quantity, which can come
back to the form of comoving curvature perturbation in the limit
where the physical wave-length is large enough to ignore nonminimal
coupling corrections.

In our model, the nearly conserved quantity takes the form
\begin{equation}\label{R}
\mathcal{R}_k\equiv \frac{H}{\dot{H}} \dot{\phi}_k+\left(
\frac{H^2}{\dot{H}}+2\frac{\dot{F}}{F}\frac{H}{\dot{H}}+\frac{\dot{G_0}}{1+G_0}\frac{H}{\dot{H}}-1
\right)\phi_k
\end{equation}

It can be examined that $\dot{\mathcal{R}_k}/(H\mathcal{R}_k)\sim
\mathcal{O}(\epsilon^2)$. The detailed derivation of this quantity
is presented in Appendix A. The issues on entropy perturbation and
the differences between the comoving curvature perturbation and the
newly-derived quantity are discussed in Appendix B. Since the nearly
conserved quantity turns smoothly into the comoving curvature
perturbation at late time during inflation, the observables like the
temperature fluctuations on the large-angular scales can be directly
related to this quantity. The amplitude of entropy perturbation
automatically decays well outside the horizon, so it does not
conflict with the experiments on the observed limits of entropy
perturbations.

\section{Quantum Theory of Perturbations}

Quantization plays an essential role in creating the perturbations
from an initial ``vacuum'' state, and sets the initial conditions
for the solutions to the perturbation equations. To do quantization,
we have to expand the action up to second order in the perturbation
variables:
\begin{equation}
\delta_2 S=\frac{1}{2}\int \left((v')^2-(\partial_i
v)^2+\frac{z{''}}{z}v^2 \right)d^4 x
\end{equation}
As usual the commutation relations on the constant conformal
time hypersurface are satified by  the canonical coordinate $v$ and its
conjugate momentum.

The equation of motion of $v$ takes the form:
\begin{equation}\label{coord}
v{''}_k+k^2v_k+m^2 v_k=0
\end{equation}

At large $k/(aH)$, provided that the perturbation is inside the
horizon and the nonminimal coupling effect has not been so strong to
spoil the local flat space quantum field theory, the normalization
can be determined as in the flat spacetime as
\begin{equation}
v_k=\frac{1}{\sqrt{2k}}e^{-ik\tau}
\end{equation}

Now recall $u$ in the pervious section, we shall determine the role
of $u$ in this quantum system. Note that $u$ can be expressed in
linear combination of $v$ and $v'$ as:
\begin{equation}\label{uv}
u_k =A v'_k + B v_k
\end{equation}

In general, $A$ and $B$ are $k$-dependent. But up to a
time-independent overall factor, they should be almost
$k$-independent. This is because the ratio of $A$ and $B$ is
determined by dynamics, and during the slow-roll inflation, each
mode has almost the same dynamics in its production and evolution.

Insert (\ref{uv}) into (\ref{momentum}), and use (\ref{coord}) to
simplify the expression. By collecting the $k^2 v$ term in the
resulting equation, we get $A'=0$, i.e. $A$ is independent of time.
Then we can determine $A$ at very late time during inflation. By
comparing the formula of nonminimal inflation model
\begin{equation}
\phi_k=\frac{A \dot{\varphi}}{F^{\frac{3}{2}}(1+G_0)}
{v_k}'+\frac{B\dot{\varphi}}{F^{\frac{3}{2}}(1+G_0)} v_k
\end{equation}
with the corresponding formula of the minimal model at $k/(aH)\ll
1$, $A$ is determined to be
\begin{equation}
A=\frac{M_p}{k^2}
\end{equation}

The validity of this equation can be checked numerically. In the
models discussed in the following sections, at a few e-folds before
the end of inflation, the energy scale is indeed low enough where
the non-minimal coupling models dynamically reduces to the minimal
model.

Having $A$ in hand, we are able to fix the initial condition of $u$.
When $k/(aH)$ is large, $A v'_k$ term in (\ref{uv}) is the main
contribution. So the short-wave limit of $u$ reads
\begin{equation}
u_k=\frac{-iM_p}{k\sqrt{2k}}e^{-ik\tau}
\end{equation}

By expanding the Bessel function at short-wave limit, one can fix
the integration constants in (\ref{usolution})
\begin{equation} \left| C_1 \right|=\sqrt{\frac{\pi}{2}} \frac{M_p}{2k} , ~~~~~~C_2=0
\end{equation}

\section{The Power Spectrum and the Spectral Index}
We now have all the pieces needed to calculate the power spectrum of
the nearly-conserved quantity $\mathcal{R}$. To the leading order in
slow-roll parameters, $\mathcal{R}$ can be expressed as

\begin{eqnarray}
\mathcal{R}_{k}&\simeq&\frac{H}{\dot{H}}\dot{\phi}_{k}+\frac{H^2}{\dot{H}}\phi_{k}\\
\nonumber
&\simeq&-\frac{M_p}{\sqrt{2}k^{3/2}}\frac{H^2}{F^{1/2}\dot{\varphi}}
\left(\frac{-k\tau}{2}\right)^{-\nu+1/2}
\end{eqnarray}
We calculate $\mathcal{R}$ at a few e-folds outside the horizon,
where the Bessel function solution still holds and $\mathcal{R}$ can
be taken as a conserved quantity from then on. The coefficients in
the Bessel equation are calculated at $k=aH$.

The power spectrum of $\mathcal{R}$ takes the form
\begin{equation}
\mathcal{P}_{\mathcal{R}}(k)\equiv\frac{k^3}{2\pi^2}\left|
\mathcal{R}_k
\right|^2=\frac{M_p^2}{4\pi^2}\frac{H^4}{F\dot{\varphi}^2}\label{spectrum}
\end{equation}
It follows that the spectral index is
\begin{equation}
n_s-1\equiv \left. \frac{d \ln \mathcal{P}_{\mathcal{R}}}{d \ln
k}\right|_{k=aH}=-4\epsilon +2\delta-\gamma\label{ns},
\end{equation}

As usual, $n_s-1$ can be calculated in another way. We can consider
the explicit $k$ dependence of $\mathcal{P}_{\mathcal{R}}$ a few
e-folds after crossing the horizon and do not use the horizon
crossing condition. Then the spectral index can be expressed by
$\nu$ in (\ref{nu}),
\begin{equation}
n_s-1=1-2\nu=-4\epsilon +2\delta-\gamma,\label{anotherns}
\end{equation}
which agrees with (\ref{ns}). It is not only a self-consistency test
for the conservation of $\mathcal{R}$, but also a test for the
quantization. Note that the only condition for (\ref{anotherns}) to
hold is that the classical evolution equation (\ref{momentum}) remains
the same outside the horizon. So in parameter regions where the
nonminimal coupling is very strong and we can not expand
$\mathcal{G}$ as equation (\ref{planewave}) inside the horizon, the
power spectrum fomula may change up to a multiplying constant, but
the spectral index (\ref{ns}) and the running of the spectral index
(\ref{alphas}) should still be correct.

The running of the spectral index is
\begin{equation}
\alpha_s\equiv \left. \frac{d n_s}{d \ln
k}\right|_{k=aH}=-\frac{4\dot{\epsilon}}{H}
+\frac{2\dot{\delta}}{H}-\frac{\dot{\gamma}}{H}\label{alphas}
\end{equation}

For the tensor modes, the results in \cite{H0412} can be directly
generalized to our models. The spectrum of the tensor modes reads
\begin{equation}
\mathcal{P}_T=\frac{2}{\pi^2}\frac{H^2}{F},
\end{equation}
where we use the normalization such that in the limit of a minimal
inflation model, the ratio of the tensor and scalar type
perturbation reduces to $r=16\epsilon$.

The spectral index of the tensor modes is
\begin{equation}
n_T=-2\epsilon-\gamma\label{nt}
\end{equation}
In the above equation, $\epsilon$ is always positive, so the tensor
power spectrum is always red in the minimal inflation models. As to
be demonstrated in the next section, $\gamma$ is negative in some
cases thus it can help to produce a blue spectrum of the tensor mode
perturbation. This is a new effect in the nonminimal models.

The ratio of the amplitude of tensor fluctuations to scalar
fluctuations is calculated to be
\begin{equation}\label{rr}
r\equiv\frac{\mathcal{P}_T}{\mathcal{P}_{\mathcal{R}}}=8\frac{F}{M_p^2}\left(2\epsilon+\gamma\right)
\end{equation}

The consistency relation has been modified to be
\begin{equation}
r=-8\frac{F}{M_p^2}n_T
\end{equation}
This new consistency relation is subject to test in future experiments.

\section{Some Nonminimal Inflation Models}
In this section, we shall examine some concrete models of the following
form
\begin{equation}
f=M_p^2 R + g(\varphi) R^{n_3}
\end{equation}
During inflation, the Hubble scale is much lower than the Planck
scale, the coupling of the inflaton with very large power of R is
unlikely to be important. We will consider the cases when $n_3=1,2$
and study the $n_3=2$ case in more detail, the latter has not been
studied in the literature except in \cite{miaoli} as far as we know,
and is the most natural candidate to make a large correction to the
minimal inflation model. Of course other choices are interesting in
their own right, although we are not going to investigate them in
this paper.

For the inflation potential $V(\varphi)$, we consider the following
potentials for the inflaton: the quadratic potential
$V(\varphi)=\frac{1}{2}m^2 \varphi^2$, the quartic potential
$V(\varphi)=\frac{1}{4}\lambda\varphi^4$, the power-law potential
$V(\varphi)=\lambda M_p^4 \exp(-c\varphi/M_p)$, the small field
quadratic potential $V(\varphi)=\lambda M_p^4 (1-c\varphi^2/M_p^2)$,
and the small field quartic potential$V(\varphi)=\lambda M_p^4
(1-c\varphi^4/M_p^4)$. Again, the results can be generalized
directly to other kinds of potentials of $\varphi$.

In our calculations, the co-moving wave number $k$ and the
nonminimal coupling constant $\lambda_0$ are taken as either input
parameters or abscissas in diagrams, while the coupling constant in
the inflaton potential $V(\varphi)$ is determined by COBE
normalization
$\delta_H^2=\frac{4}{25}\mathcal{P}_{\mathcal{R}}\approx 4\times
10^{-10}$.

\subsection{Models with $f=M_p^2 R + \lambda_0 \varphi^2 R$}
As we know, this class of models can be related to a minimal
inflation model with a single inflaton by a conformal
transformation, and this kind of coupling is not of particular
interest in quantum gravity theories such as string theory. But to
compare with the models listed below, we shall still discuss some
features of this class of models.

\begin{figure}
\centering
\includegraphics[totalheight=2.4in]{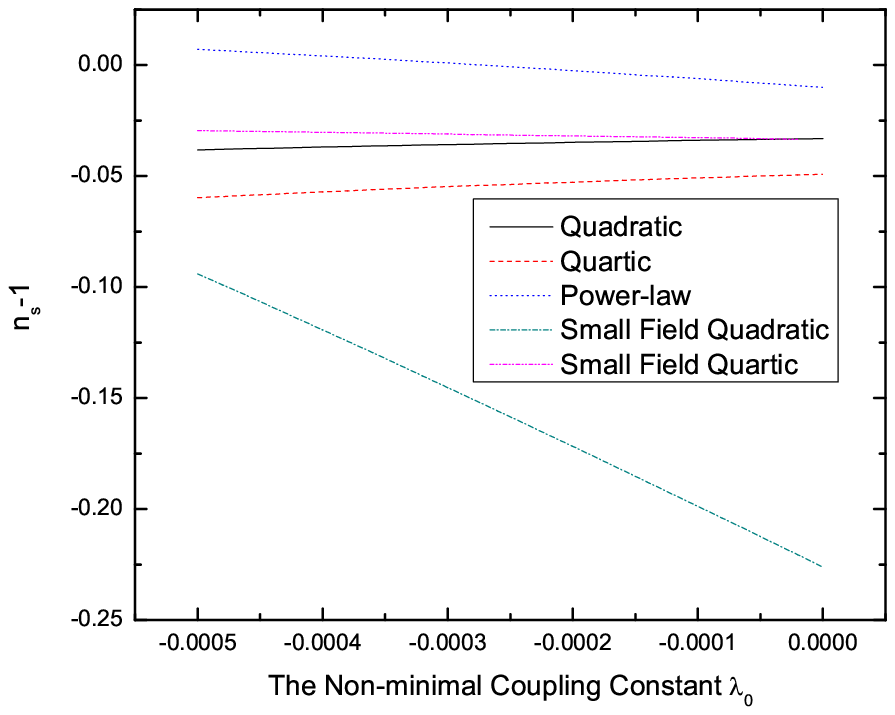}
\includegraphics[totalheight=2.4in]{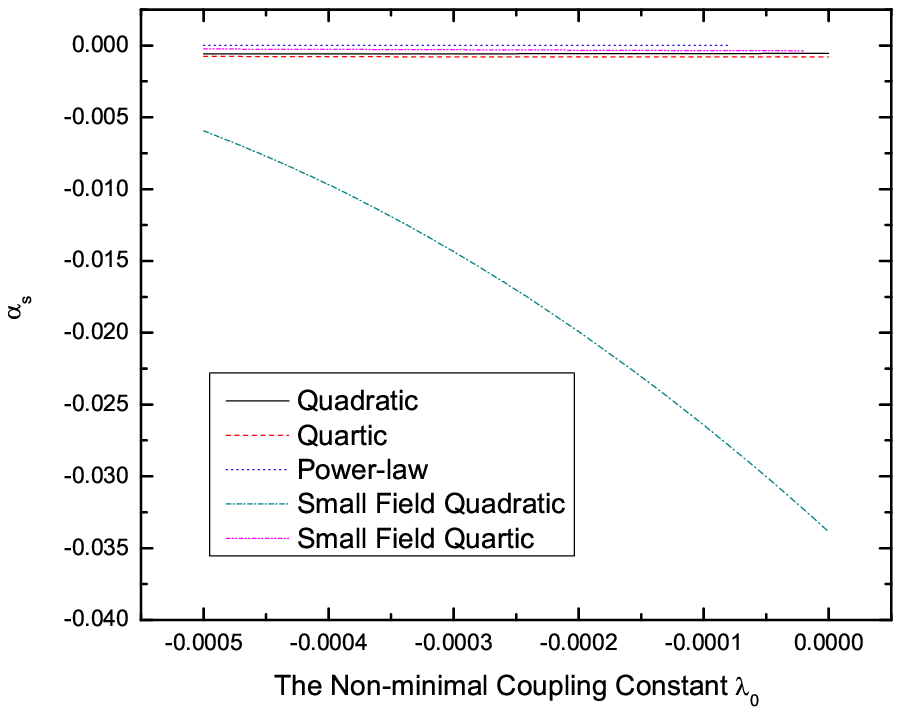}
\caption{\small{$n_s-1$ and $\alpha_s$ as functions of $\lambda_0$
in the models with $f=M_p^2 R + \lambda_0 \varphi^2 R$. Red spectrum
is common in this case except that the power-law potential gives a
very small positive $n_s-1$. Running of spectral indices are quite
small except in the small field quadratic inflation.}}
\label{fig:2x1ns}
\end{figure}

\begin{figure}
\centering
\includegraphics[totalheight=2.4in]{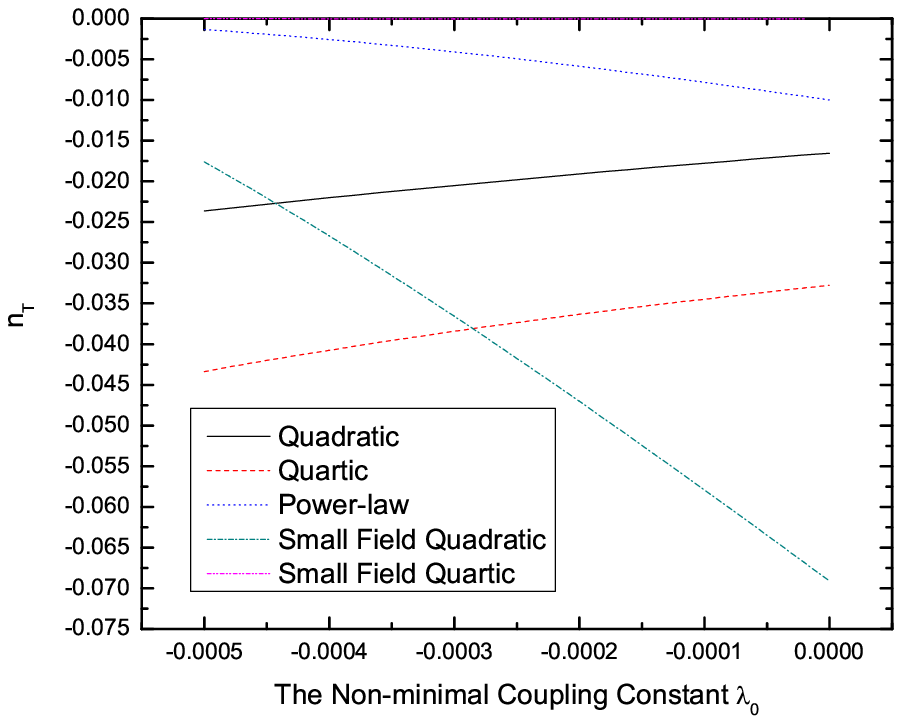}
\includegraphics[totalheight=2.4in]{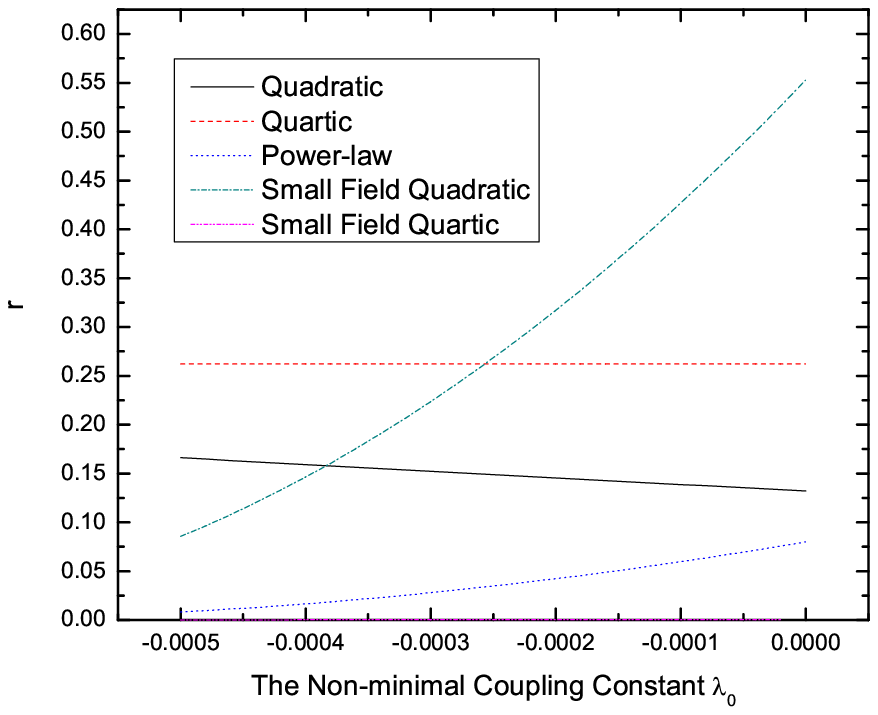}
\caption{\small{$n_T$ and $r$ as functions of $\lambda_0$ in the
models with $f=M_p^2 R + \lambda_0 \varphi^2 R$. }}
\label{fig:2x1nt}
\end{figure}

Fig.\ref{fig:2x1ns} and Fig.\ref{fig:2x1nt} present $n_s-1$,
$\alpha_s$, and $n_T$, $r$ as functions of the nonminimal coupling
constant respectively. These models generally produce a red power
spectrum, and a very small running. The power-law potential is an
exception, which is to be discussed in more detail in the next
subsection.

\subsection{Models with $f=M_p^2 R + \lambda_0  M_p^{-2}\varphi^2 R^{2}$}

\begin{figure}
\centering
\includegraphics[totalheight=2.4in]{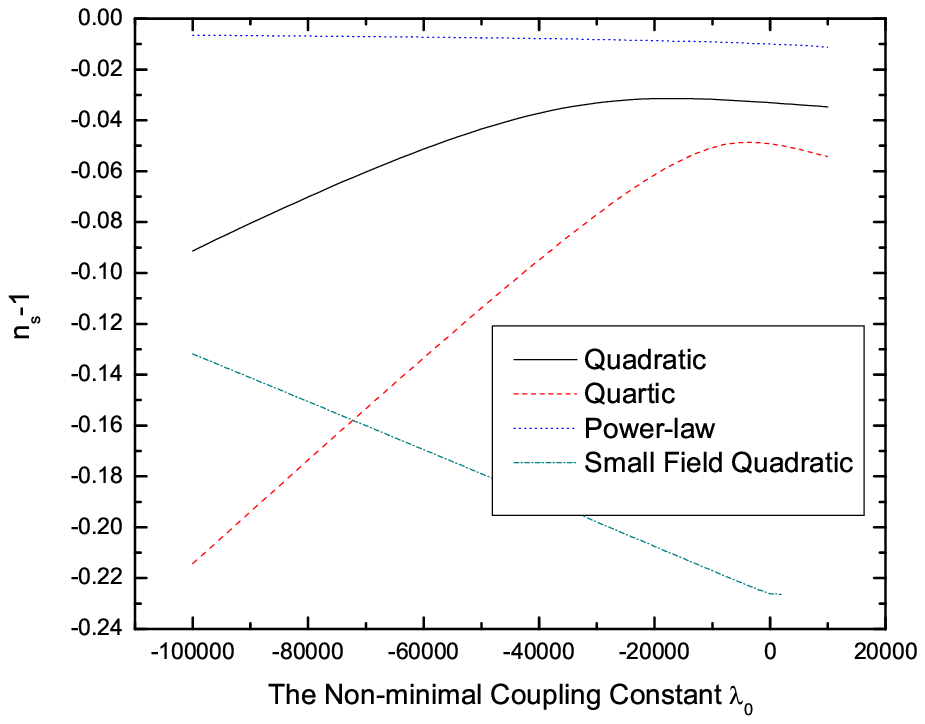}
\includegraphics[totalheight=2.4in]{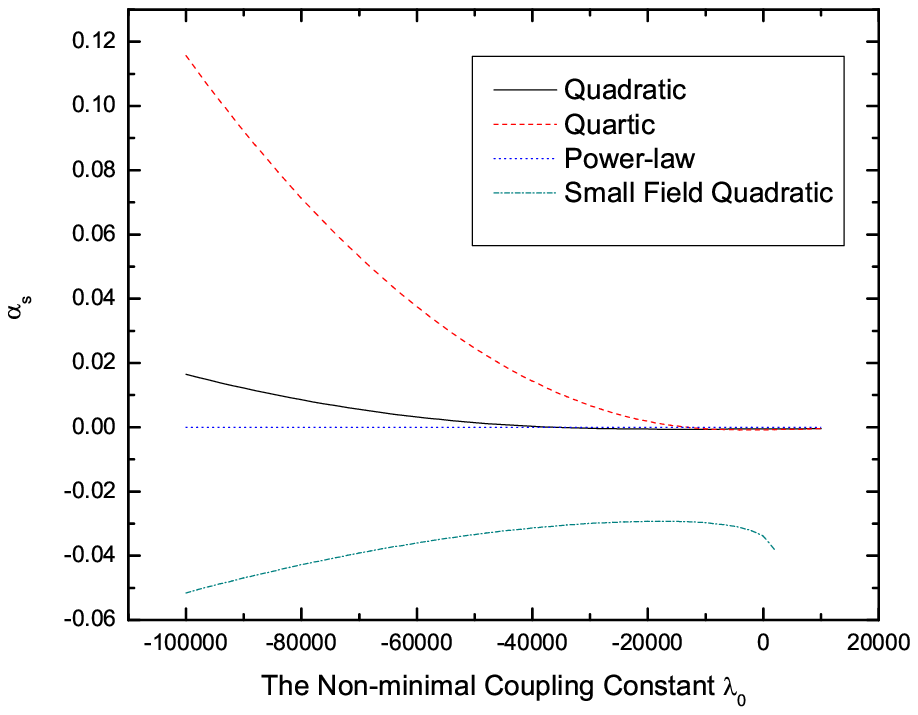}
\caption{\small{$n_s-1$ and $\alpha_s$ as functions of $\lambda_0$
in the models with $f=M_p^2 R + \lambda_0  M_p^{-2}\varphi^2 R^{2}$
for the solution which has a smooth minimal coupling limit. The
power spectrum is also red in all cases we consider. Large running
can be produced when the absolute value of $\lambda_0$ is large, but
only the small field quadratic potential produces a negative large
running.}} \label{fig:2x2ns}
\end{figure}

\begin{figure}
\centering
\includegraphics[totalheight=2.4in]{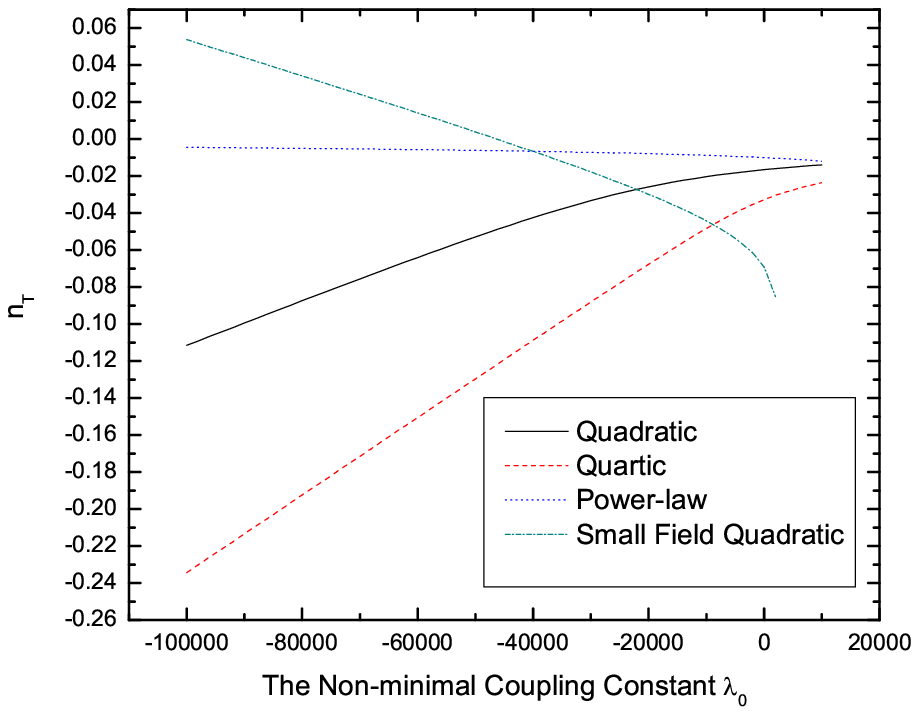}
\includegraphics[totalheight=2.4in]{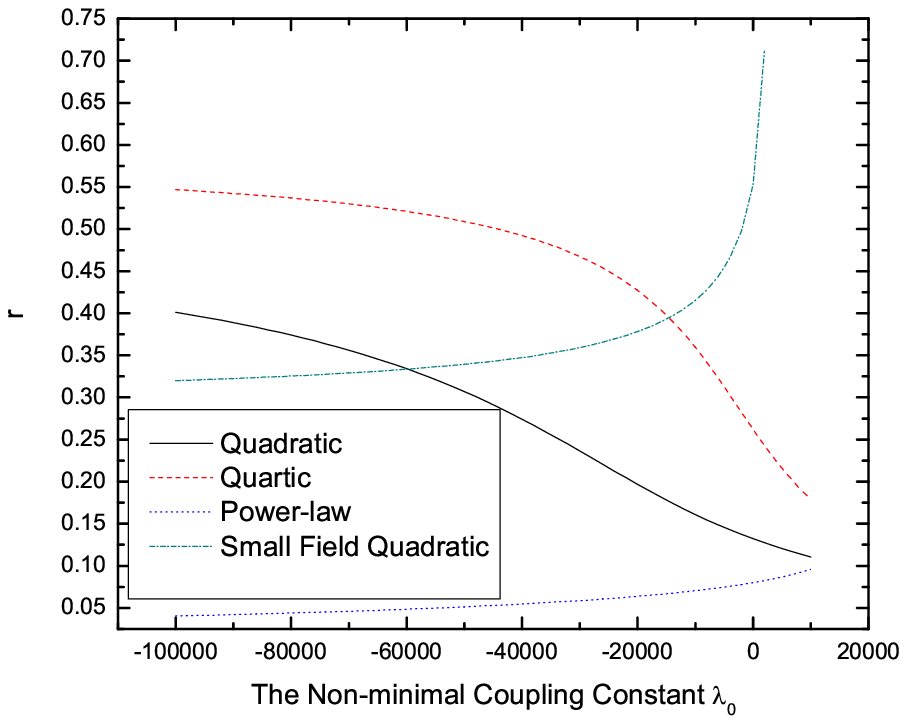}
\caption{\small{$n_T$ and $r$ as functions of $\lambda_0$ in the
models with $f=M_p^2 R + \lambda_0 M_p^{-2}\varphi^2 R^{2}$ for the
solution which has a smooth minimal coupling limit. The small field
quadratic potential model provide a positive $n_T$. Such phenomena
never appears in the minimal models.}} \label{fig:2x2nt}
\end{figure}

By COBE normalization condition and choosing the solution which
smoothly reduces to the minimal case in the weak nonminimal coupling
limit, $n_s-1$, $\alpha_s$, and $n_T$, $r$ are shown in
Fig.\ref{fig:2x2ns} and Fig.\ref{fig:2x2nt} respectively. A red
power spectrum is obtained in this case. A large running can be
produced for some potentials when the nonminimal coupling is large.
For the small field potential the large running is negative while
the quadratic potential gives a positive large running.

We note from these figures that the small field inflation with
$V(\varphi)=\lambda(1-c\varphi^2)$ is not favored by experiment when
$\lambda_0\rightarrow 0$ because $n_s-1$ is too small and $r$ is too
large. But when $\lambda_0$ is taken into consideration, this model
behaves much better in both figures. We also observe that $n_T$ in
the small field inflation can be positive, while in the minimal
inflation model it is always negative. If in future experiments a
blue spectrum of gravitational waves is detected, it may be a signal
that nonminimal coupling effect should be considered.

\begin{figure}
\centering
\includegraphics[totalheight=2.4in]{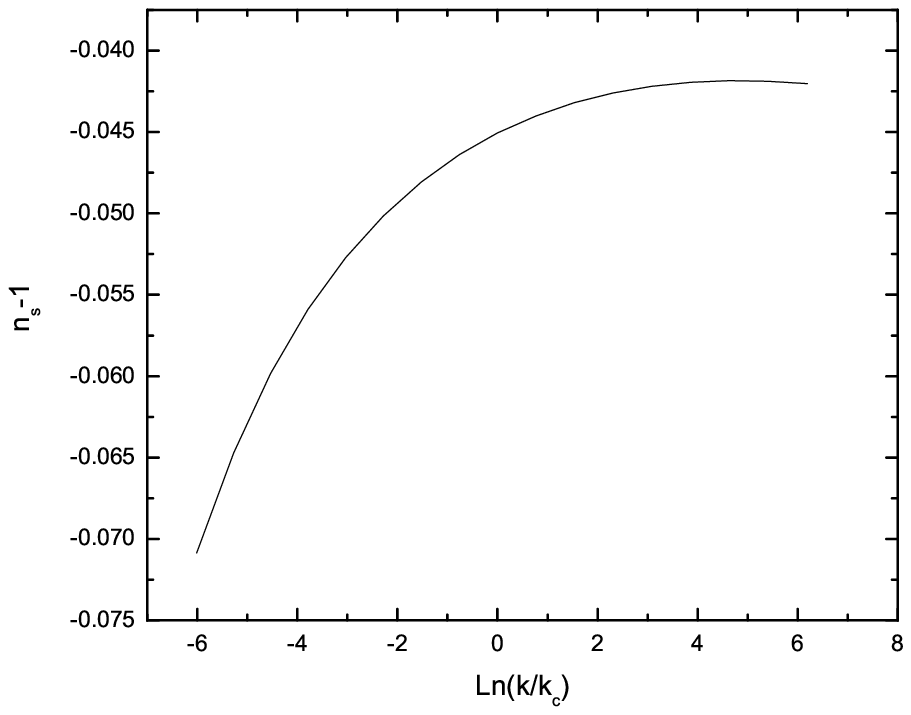}
\includegraphics[totalheight=2.4in]{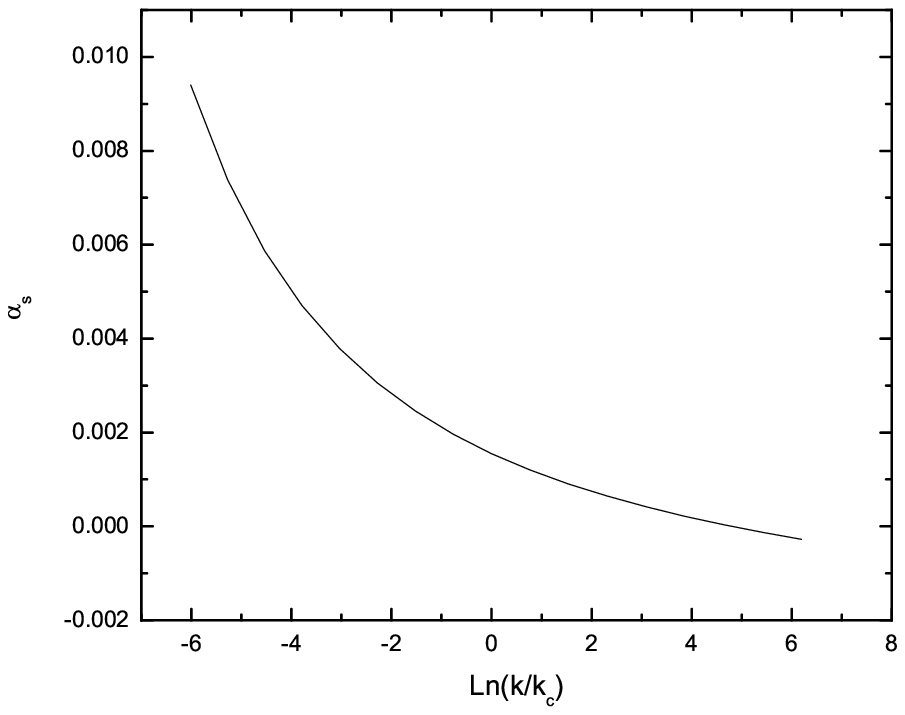}
\caption{\small{$n_s-1$ and $\alpha_s$ as functions of $\ln (k/k_c)$
in the models with $f=M_p^2 R + \lambda_0  M_p^{-2}\varphi^2 R^{2}$
and the quadratic potential $V(\varphi)=\frac{1}{2}m^2 \varphi^2$
for the solution which has a smooth minimal coupling limit.
$\lambda_0$ is taken -50000 in both figures. $k_c$ is where the COBE
normalization is taken. We assume at $k_c$, there are 60 e-folds of
inflation. The power spectrum is red, with a positive running.}}
\label{fig:2x2nsrunk}
\end{figure}

Fig.\ref{fig:2x2nsrunk} presents $n_s-1$ and $\alpha_s$ as a
function of $\ln(k/k_c)$, where $k_c$ is where COBE normalization is
taken, we assume it is 60 e-folds from the end of inflation. To know
the exact e-folding number at COBE normalization point requires the
knowledge of the details of reheating, which will not be considered
in this paper.

We find that something new happens here, if we take $\lambda_0\sim
-60000$ and draw a figure like Fig.\ref{fig:2x2nsrunk}. $F$ crosses
zero at the largest scale that can be observed today, and the
perturbations blow up during the zero-crossing. This region can not
be treated exactly by the approach in this paper because the
slow-roll conditions (\ref{slowroll}) and  the expansion
(\ref{planewave}) break down. The exact equations of motion for the
background shows that this region is indeed unusual. This phenomenon
can also happen in some other models studied in this paper. This
phenomenon may have something to do with the lower CMB anisotropy
power on the largest angular scales. To make sure of this, a
detailed calculation is required. Similar phenomenon in the
quintessential models of dark energy is studied \cite{DEF}, but as
far as we know, no consideration in inflation models and effects on
the power spectrum is made.

\begin{figure}
\centering
\includegraphics[totalheight=2.4in]{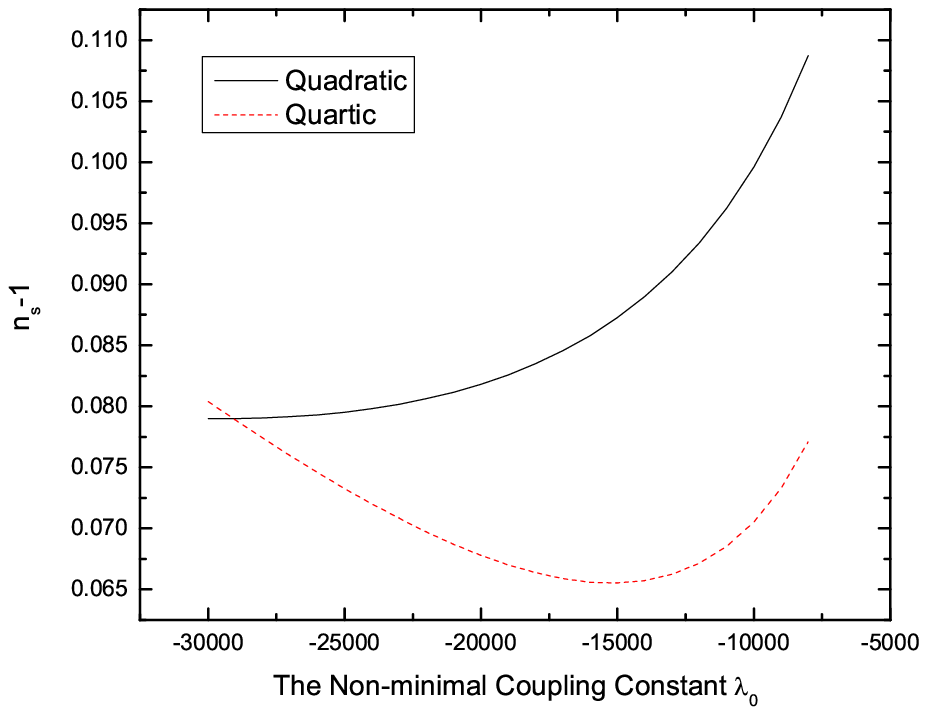}
\includegraphics[totalheight=2.4in]{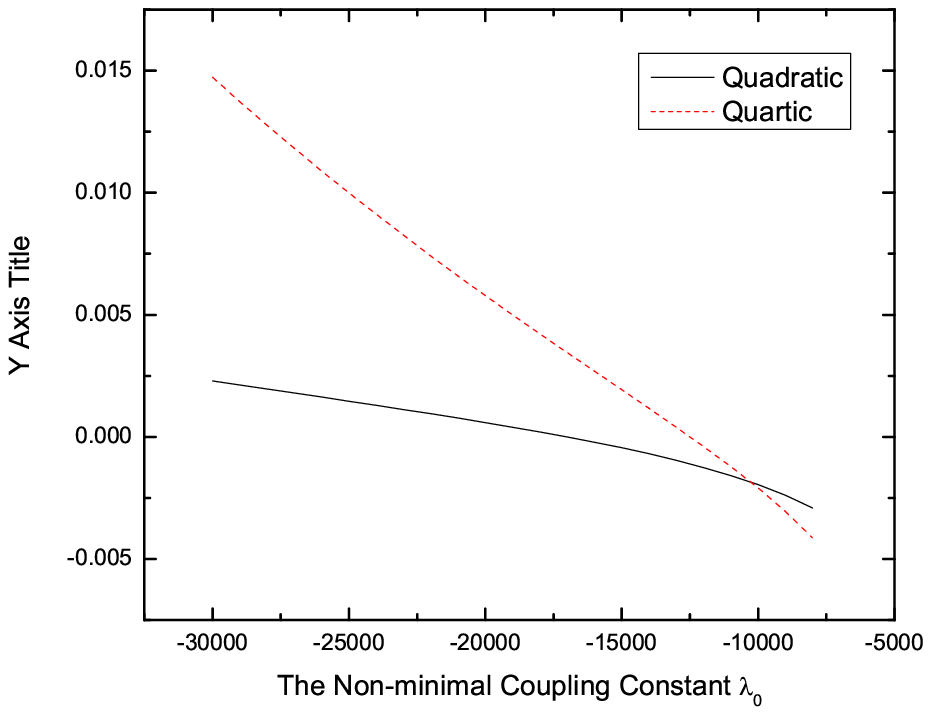}
\caption{\small{$n_s-1$ and $\alpha_s$ as functions of $\lambda_0$
in the models with $f=M_p^2 R + \lambda_0  M_p^{-2}\varphi^2 R^{2}$
for the new solution in the nonminimal model. A blue power spectrum
is produced in this case, with a negative running when the absolute
value of $\lambda_0$ is not large.}} \label{fig:another2x2ns}
\end{figure}

\begin{figure}
\centering
\includegraphics[totalheight=2.4in]{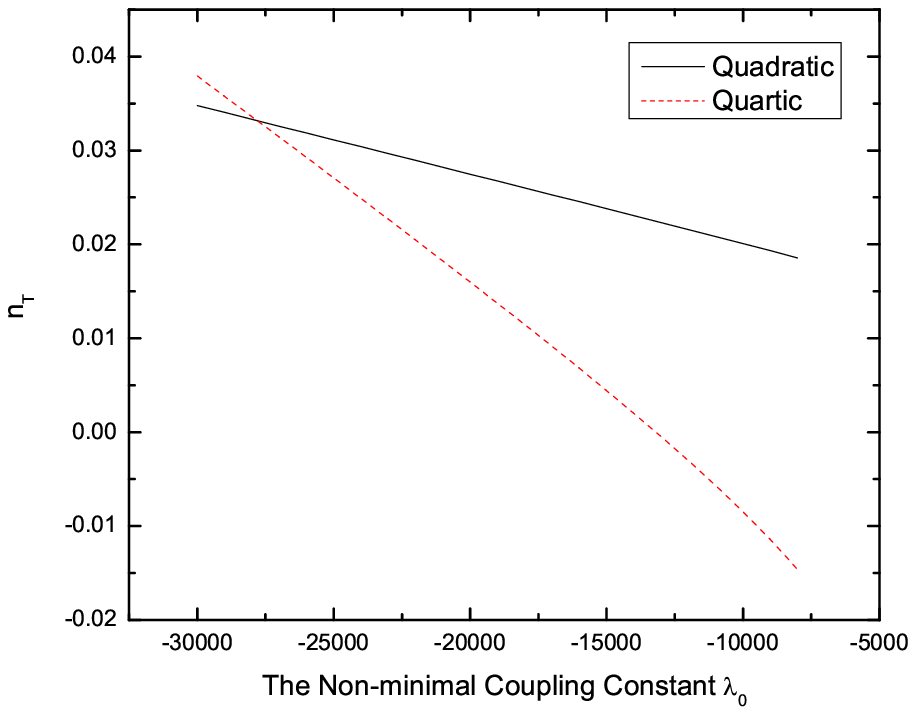}
\includegraphics[totalheight=2.4in]{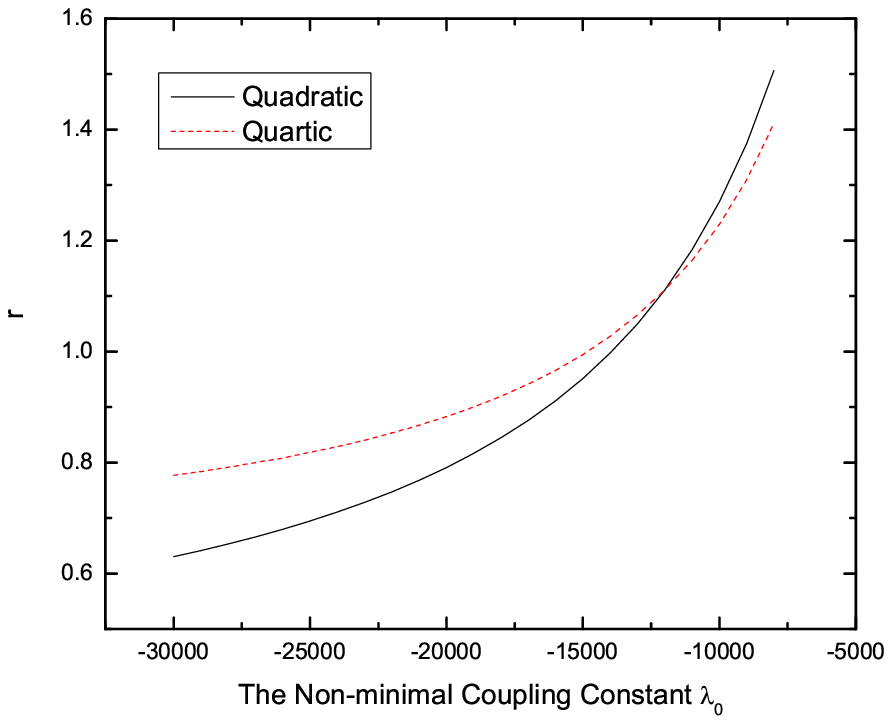}
\caption{\small{$n_T$ and $r$ as functions of $\lambda_0$ in the
models with $f=M_p^2 R + \lambda_0  M_p^{-2}\varphi^2 R^{2}$ for the
new solution in the nonminimal model. Tensor mode perturbation is
large in this solution, which may be approved or disapproved in the
near future.}} \label{fig:another2x2nt}
\end{figure}

The COBE normalization equation does not always have an unique
solution. In some models, there can be additional solutions other
than the solution studied above, namely, there are different regions
where either $V$ or $f$ dominates. These new solutions can be much
different from the minimal inflation models. In these models, a blue
power spectrum is common and a negative running of the spectral
index is possible. Large gravitational waves can be produced in
these models, while in a large parameter region $r$ lies below the
experimental upper bound. Models of this kind with the quadratic and
quartic potential are drawn in Fig.\ref{fig:another2x2ns} and
Fig.\ref{fig:another2x2nt}.

\begin{figure}
\centering
\includegraphics[totalheight=2.4in]{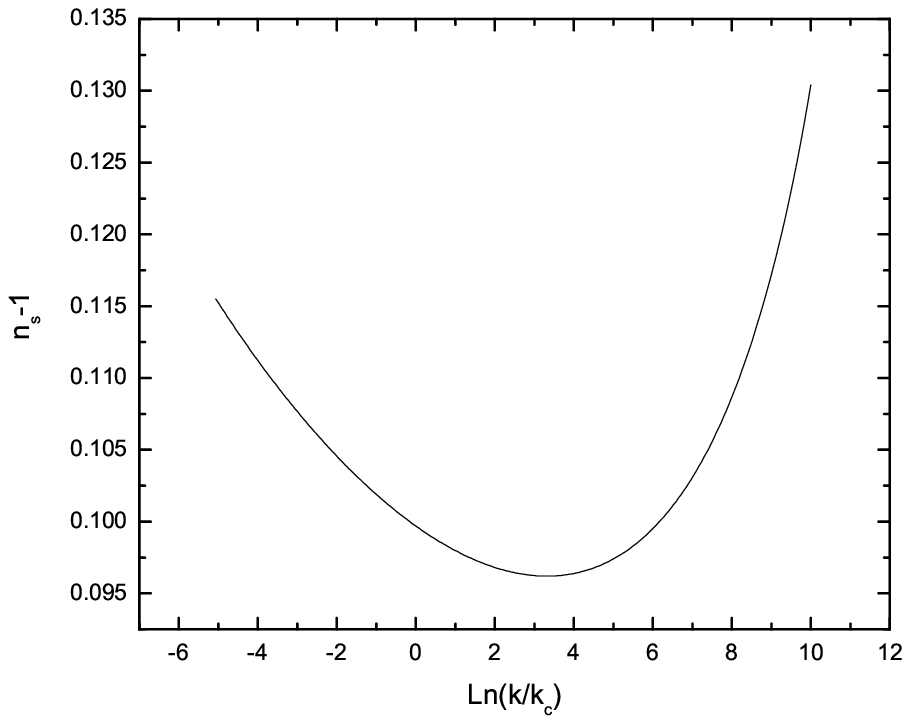}
\includegraphics[totalheight=2.4in]{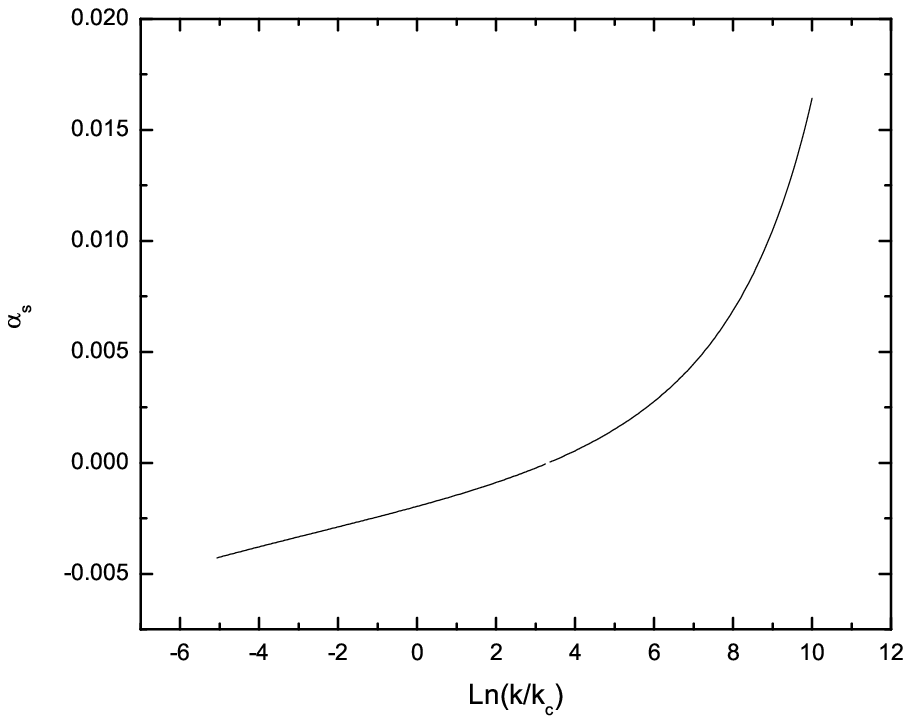}
\caption{\small{$n_s-1$ and $\alpha_s$ as functions of $\ln (k/k_c)$
in the models with $f=M_p^2 R + \lambda_0  M_p^{-2}\varphi^2 R^{2}$
and the quadratic potential $V(\varphi)=\frac{1}{2}m^2 \varphi^2$
when $f$ dominates over $V$. $\lambda_0$ is taken -10000 in both
figures. In order that $r$ is not too large, the running can not be
large enough to change the blue spectrum to a red one.}}
\label{fig:another2x2nsrunk}
\end{figure}

Fig.\ref{fig:another2x2nsrunk} presents the $k$-dependence of
$n_s-1$ and $\alpha_s$ for this solution. For the model we are
considering, the running is not large enough for the blue spectrum
to run to a red spectrum.

By comparing with the $n_3=0$ minimal models and the corresponding
figures for the models with $n_3=1$, we find that a large running is
usually produced in $n_3=2$ models. While in the $n_3=1$ and $n_3=0$
case, it is difficult to produce such a large running except in the
small field quadratic model.

It can be shown that the class of models $f=M_p^2 R + \lambda_0
M_p^{-4}\varphi^4 R^{2}$ are quite similar to
models $f=M_p^2 R + \lambda_0  M_p^{-2}\varphi^2 R^{2}$, the only
important difference is that in order to get the same $n_s-1$
and the running, $\lambda_0$ required in the former
models are commonly 3 or 4 magnitudes smaller than that in the
latter models.

\subsection{Models with $f=M_p^2 R + \lambda_0 \exp(-\lambda_1 \varphi/M_p) R^{2}$}
\begin{figure}
\centering
\includegraphics[totalheight=2.4in]{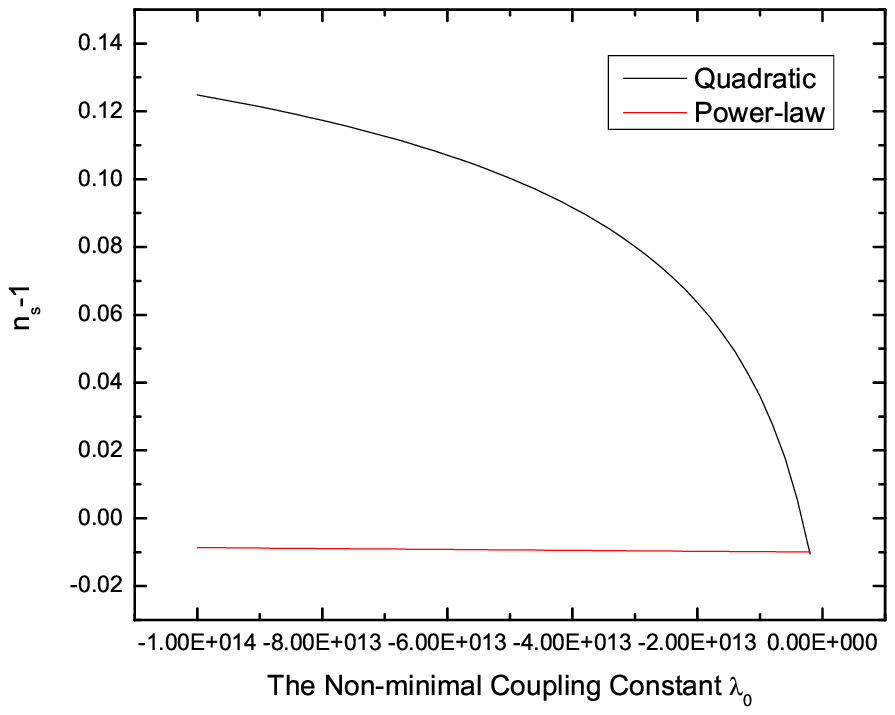}
\includegraphics[totalheight=2.4in]{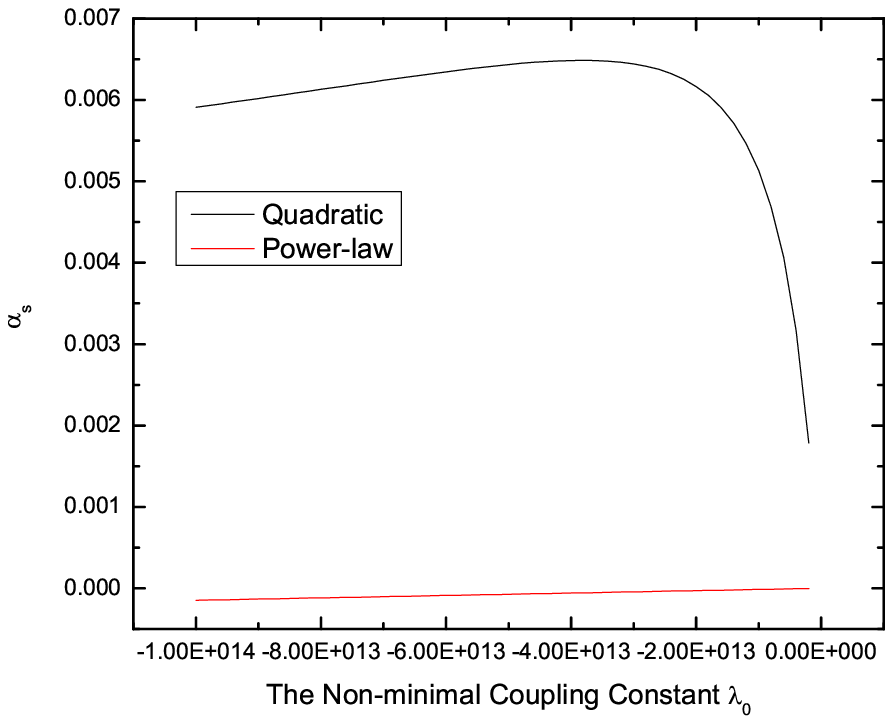}
\caption{\small{$n_s-1$ and $\alpha_s$ as  functions of $\lambda_0$
in the the models with $f=M_p^2 R + \lambda_0 \exp(-\lambda_1
\varphi) R^{2}$. A blue spectrum is produced in the quadratic
potential, but the running is positive. A number $-1.00E+014$ in the
figure denotes $-1.00\times 10^{14}$.}} \label{fig:ex2ns}
\end{figure}

\begin{figure}
\centering
\includegraphics[totalheight=2.4in]{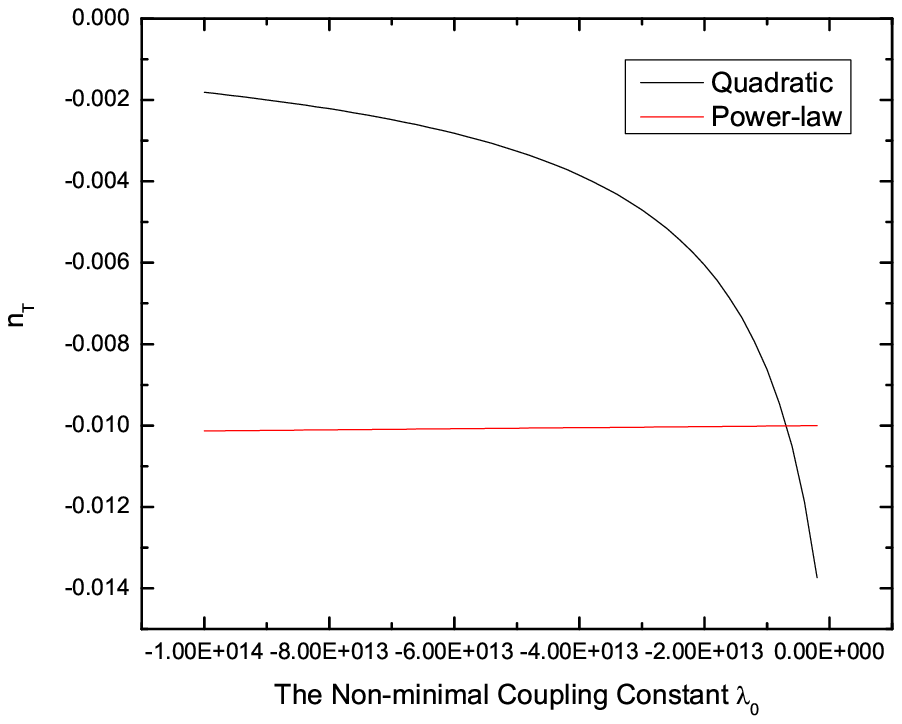}
\includegraphics[totalheight=2.4in]{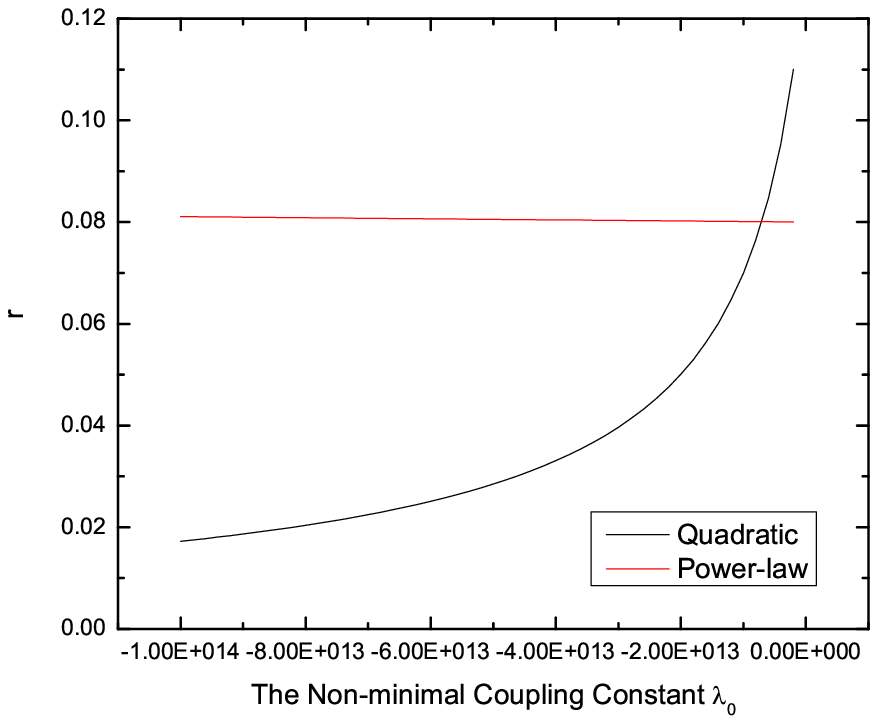}
\caption{\small{$n_T$ and $r$ as functions of $\lambda_0$ in the
models with $f=M_p^2 R + \lambda_0 \exp(-\lambda_1 \varphi)
R^{2}$.}} \label{fig:ex2nt}
\end{figure}

In this class of models, to make any differences from the minimal
model, a very large $\lambda_0$ is required. A blue power spectrum
can be produced with a large but positive running. $n_s-1$,
$\alpha_s$, and $n_T$, $r$ are shown in Fig.\ref{fig:ex2ns} and
Fig.\ref{fig:ex2nt} respectively.

\section{Conclusion}

We obtained  the scalar power spectrum $\mathcal{P}_{\mathcal{R}}$,
the spectral index $n_s-1$, the running of the spectral index
$\alpha_s$, the tensor mode spectral index $n_T$, and the ratio of
the amplitude of tensor fluctuations to scalar fluctuations $r$ in
this paper, they  are given in equations (\ref{spectrum}),
(\ref{ns}), (\ref{alphas}), (\ref{nt}) and (\ref{rr}) respectively.
We constructed a nearly-conserved quantity which can be used in
models different from the single inflaton model with minimal
coupling. It has a wide range of applicability. Also, a new
consistency relation is given, which is subject to test in future
experiments.

In the study of some concrete models, we see that there can be
novel new features, such as a blue spectrum and a large running.
We can also arrange  $F$ properly in order to have a brief
fast-rolling before the slow-rolling period. It has been shown
that the problem of a large running of the spectral index is eased
in our models in contrast to the minimal model or $f(\varphi) R$
coupling models.

The following issues deserve further investigation: The case when a
nonminimal coupling becomes so strong that the quantum gravity
effect destroys the usual canonical quantization picture at a length
scale much smaller than the Hubble radius; the region where $F$
crosses zero, producing a sudden fast-roll stage during the period
of slow-roll inflation. A more careful study of these issues may
lead to very interesting physics.

\section*{Acknowledgments}

This work was supported by grants of NSFC. BC was also supported by
the Key Grant Project of Chinese Ministry of Education (NO. 305001).
We are grateful to Jianxin Lu for reading the nmanuscript.

\appendix

\section{A General Nearly Conserved Quantity}
From a general differential equation with two small parameters
$\epsilon_1$ and $\epsilon_2$,
\begin{equation}
\label{generaldiff} \ddot{\phi}+(1+\epsilon_1)H\dot{\phi}+\epsilon_2
H^2\phi+\frac{q^2}{a^2}\phi=0
\end{equation}
The following quantity is nearly conserved when $q\ll aH$,
\begin{equation}
\mathcal{R}=c[\frac{1}{H}\dot{\phi}+(1+\epsilon_1+\frac{\dot{H}}{H^2}-\epsilon_2)\phi]e^{\int\epsilon_2
Hdt}
\end{equation}
where c is a constant. $\dot{\mathcal{R}}/(H\mathcal{R})$ is of
order $\mathcal{O}(\epsilon^2)$ plus a term proportional to
$q^2/a^2$.

\begin{equation}
\mathcal{\dot{R}}
=c[\epsilon_2(\epsilon_1+\frac{\dot{H}}{H^2}-\epsilon_2)H+\dot{\epsilon_1}+\left(\frac{\dot{H}}{H^2}\right)^{\cdot}-\dot{\epsilon_2}]\phi
e^{\int\epsilon_2 Hdt}-\frac{c}{H}\frac{q^2}{a^2} e^{\int\epsilon_2
Hdt}
\end{equation}

This construction is unique up to the constant $c$, and can be used
in various cases where (\ref{generaldiff}) is satisfied.

In the minimal inflation model, the equation (\ref{generaldiff})
takes the form
\begin{equation}
\ddot{\phi}+(H-\frac{\ddot{H}}{\dot{H}})\dot{\phi}+(2\dot{H}-\frac{H\ddot{H}}{\dot{H}})\phi+\frac{k^2}{a^2}\phi=0
\end{equation}
And $\mathcal{R}$ is the comoving curvature perturbation, which
happens to be conserved to all orders in $\epsilon$.

In our nonminimal model, Inserting the Taylor expansion of ${\cal
G}_k$ into (\ref{phidiag}), one can find
\begin{eqnarray}
&&\phi{''}_k+(aH)\left(\frac{2\dot{G}_0}{H(1+G_0)}+2\delta+3\gamma\right)\phi'_k+q^2\phi_k \nonumber\\
&&+(aH)^2\left(\frac{\dot{G_0}}{H(1+G_0)}-\frac{4G_0}{1+G_0}\epsilon
+\frac{3+G_0}{1+G_0}\gamma+2\delta \right)\phi_k=0 ,
\end{eqnarray}
where $q$ is a rescale of $k$ as in (\ref{rescale}). So
$\mathcal{R}$ can be written as
\begin{equation}
\mathcal{R}_k\equiv \frac{H}{\dot{H}} \dot{\phi}_k+\left(
\frac{H^2}{\dot{H}}+2\frac{\dot{F}}{F}\frac{H}{\dot{H}}+\frac{\dot{G_0}}{1+G_0}\frac{H}{\dot{H}}-1
\right)\phi_k
\end{equation}

It is shown in Appendix B that calculations using the comoving
curvature perturbation can also get correct results in our
nonminimal model. But it can be shown that this fact is not generic
in other generalizations of inflation models.

The only requirement for $\mathcal{R}$ to be nearly conserved is a
diagonized equation of motion. So it can also be used in many other
cases such as inflation models with holographic dark energy
components \cite{miaoliDE} where entropy perturbation is produced and
the comoving curvature perturbation is not a  conserved
quantity.

\section{Entropy-Perturbation-Like Sources}

In the minimal inflation model, during inflation and the
post-inflationary period before the comoving wave length re-enter
the horizon, the pressure can be expressed as $p=p(\rho, S)$, where
$\rho$ is the energy density and $S$ is the entropy density up to a
normalization. If there is no entropy perturbation produced, just as
in the single-field minimal inflation model, there is only one
degree of freedom in the scalar type perturbations, so $p$ only
depends on $\rho$. Then one can find
\begin{equation}
\frac{\dot{\rho}}{\dot{p}}=\frac{\delta \rho}{\delta p}
\label{adiabatic}
\end{equation}

This relation has two consequences. First, when (\ref{adiabatic}) is
satisfied, the equations (\ref{purt1}) and (\ref{purt2}) can be
diagonized into a one-variable differential equation exactly, as in
the $f(R)$ or $f(\varphi)R$ gravity models, studied in \cite{H0412}.
Second, the comoving curvature perturbation is conserved in a model
without entropy perturbation.

But in general, when entropy perturbation $\delta S$ is generated,
the entropy perturbation can be defined as
\begin{equation}
\delta p=\frac{\dot{p}}{\dot{\rho}}\delta\rho+T\delta S
\end{equation}
where $T$ is the temperature. It can be shown that
\begin{equation}
\dot{\tilde{\mathcal{R}}}=-\frac{H}{\dot{H}}\frac{T}{2M_p^2}\delta
S+ \mbox{terms proportional to} \frac{k^2}{a^2}
\end{equation}
where $\tilde{\mathcal{R}}$ is the comoving curvature perturbation.

During inflation, $\tilde{\mathcal{R}}$ can be expressed as
\begin{equation}
\tilde{\mathcal{R}}=\psi-\frac{H}{\dot{H}}\left(\dot{\psi}+H\phi\right)
\end{equation}

In nonminimal models, for general $f(\varphi,R)$, it is different
from $\mathcal{R}$ introduced in Appendix A. From the uniqueness of
$\mathcal{R}$, we conclude that $\delta S$ is nonzero. The existence
of $\delta S$ can also be shown by direct calculations.

It is not surprising that the entropy perturbation is produced in
our nonminimal models, because when we consider nonminimal models
other than $f(R)$ or $f(\varphi)R$, one extra low energy effective
degree of freedom opens up, and the entropy perturbation can be
produced.

Although the entropy perturbation in our models exists and affects
the evolution of scalar type perturbations, it can not live so long
to have direct observable effects. To see this, we consider the
period begin from a few e-folds outside the horizon, where the long
wave length expansion can be applied to the Bessel function
solution. During this period, $\psi$ and $\phi$ can be considered as
slow-roll quantities, and the difference between time derivatives of
$\mathcal{R}$ and $\mathcal{\tilde{R}}$ can be neglected. So
\begin{equation}
\frac{\dot{\tilde{\mathcal{R}}}}{H\tilde{\mathcal{R}}}=\mathcal{O}(\epsilon^2),
\end{equation}
which means the entropy perturbation can be neglected a few e-folds
after horizon crossing, and do not destroy the observational bounds.
No extra decay mechanism for $\delta S$ is needed.

\end{document}